\documentclass[11pt,twoside]{article}
\usepackage{hepro5}
\usepackage{graphicx}

\usepackage[T1]{fontenc} % Computer Modern (CM) fonts

\usepackage{latexsym}
\usepackage{verbatim}

\def\ltsima{$\; \buildrel < \over \sim \;$}
\def\simlt{\lower.5ex\hbox{\ltsima}} % < over ~
\def\gtsima{$\; \buildrel > \over \sim \;$}
\def\simgt{\lower.5ex\hbox{\gtsima}} % > over ~

\begin{document}

\vskip 1.0cm
\markboth{E.~Pian}{Overview of relativistic jets}
\pagestyle{myheadings}

\vspace*{0.5cm}
\title{Relativistic jets: an overview of recent progress}

\author{Elena~Pian$^{1,2}$}
\affil{$^1$INAF IASF Bologna, Via P. Gobetti 101, 40129 Bologna, Italy\\
$^2$Piazza dei Cavalieri 7, 56126 Pisa, Italy}

\begin{abstract}
Despite their different nature and physics, blazars and gamma-ray bursts have in common very powerful relativistic jets, which make them the most luminous sources in the Universe.  The energy extraction from the central compact object, the jet collimation, the role and geometry of the magnetic fields, the structure of the jet itself represent still big enough questions that a complete paradigm cannot yet be drawn.   
This article is concerned with the main observational facts about blazars and gamma-ray burst jets, based on  multi-wavelength campaigns, and on the clues one can glean from these on jet formation, behavior and powering.    The future generation of telescopes and instruments and the contributions from multi-messenger investigation (astroparticle diagnostics and gravitational waves) will warrant further significant progress.

\end{abstract}

\section{Introduction}

Jets are ubiquitous in astrophysics, both in the Galaxy  from parsec  (protostars) to kilo-parsec   (X-ray binaries, a.k.a. micro-quasars) scales, and outside, in nearby radio-galaxies like Centaurus A  (4 Mpc) and M87 (15 Mpc), and at cosmological distances, where they are responsible for the  blazar and gamma-ray burst (GRB) phenomenon.  Accordingly, their observed luminosities span a wide range and the kinematic regimes of the expanding plasma  are very different, from Newtonian in protostellar objects, to ultra-relativistic in GRBs.  In X-ray binaries, where the bolometric luminosities reach $\sim 10^{39}$ erg~s$^{-1}$ (Mirabel \& Rodr{\'{\i}}guez  1999; Fender \& Belloni 2004), Lorentz factors of a few can be directly estimated from observation of both jet and counter-jet.
Blazars can reach luminosities of $\sim 10^{48}$ erg~s$^{-1}$ during outbursts, when generally the gamma-ray output dominates the total observed luminosity, and Lorentz factors of $\sim$10-20 are inferred from superluminal motions (Jorstad et al. 2013) and compactness arguments (McBreen 1979; Maraschi, Ghisellini, \& Celotti  1992).   Gamma-ray luminosities of $\sim 10^{51}$ erg~s$^{-1}$ and energy outputs of $\sim 10^{52}$ erg are observed  in GRBs (e.g. Amati et al. 2008), that are the most extreme macroscopic relativistic sources in the Universe, with Lorentz factors in  excess of 100 (M{\'e}sz{\'a}ros 2002; Piran 2004).   

According to a now widely accepted unifying scenario (Urry \& Padovani 1995), blazar jets are the analogues of the  kilo-parsec elongated structures angularly resolved with the VLBI in radio-galaxies, only closely (less than 10 degrees) aligned to the line of sight. On the other hand, in GRBs, while there is evidence of plasma relativistic expansion  from radio observations of the  afterglows, that has made possible the derivation of a Lorentz factor at a few days after the explosion (Frail et al. 1997; Mesler et al. 2012), no direct observation of jets has ever been reported.  However, their presence is inferred from the model-independent argument that the huge observed energy outputs (on average $10^{52}$ erg and sometimes in excess of $10^{54}$ erg) coupled with the millisecond variability timescales imply that the binding energy of a stellar-size collapsing/exploding object is transformed into radiation with 100\% efficiency.  This leads almost naturally to the conclusion that the radiation, instead of being emitted isotropically, must be collimated in a narrow beam, whose aperture is estimated to be of a few degrees (e.g. Frail et al. 2001; Grupe et al. 2006).   Typical signatures of jets are thought to be the achromatic temporal breaks seen in their afterglow light curves, and due to the fact that plasma deceleration and Lorentz factor drop make the jet edge causally connect with the observer  (Rhoads 1999;  Beuermann et al. 1999; Harrison et al. 1999; Israel et al. 1999; Stanek et al. 1999; Panaitescu \& Kumar 2002).

Recently, Nemmen et al. (2012) have shown that jets produced by blazars and GRBs exhibit the same correlation between the kinetic power carried by accelerated particles and the gamma-ray luminosity, with blazars and GRBs
lying at the low- and high-luminosity ends, respectively, of the correlation (Figure 1). This result implies that the efficiency of energy dissipation in jets  is similar over 10 orders
of magnitude in jet power, establishing a physical analogy between blazars and GRBs, despite the macroscopic differences, i.e. primarily the persistence of  jets in blazars, as opposed to their rapidly transient nature in GRBs.  This leads us to compare  these two classes of sources, in an attempt to clarify how jet
physics scales from compact central engines in GRBs (stellar black holes or highly magnetized rotating neutron stars) to  supermassive black holes in active galactic nuclei.  

%%%%%%%%%    FIGURE 1:  NEMMEN    %%%%%%%%%%%

\begin{figure} 
\begin{center}
%\hspace{0.25cm}
\begin{tabular}{cc}
\includegraphics[height=8.0cm]{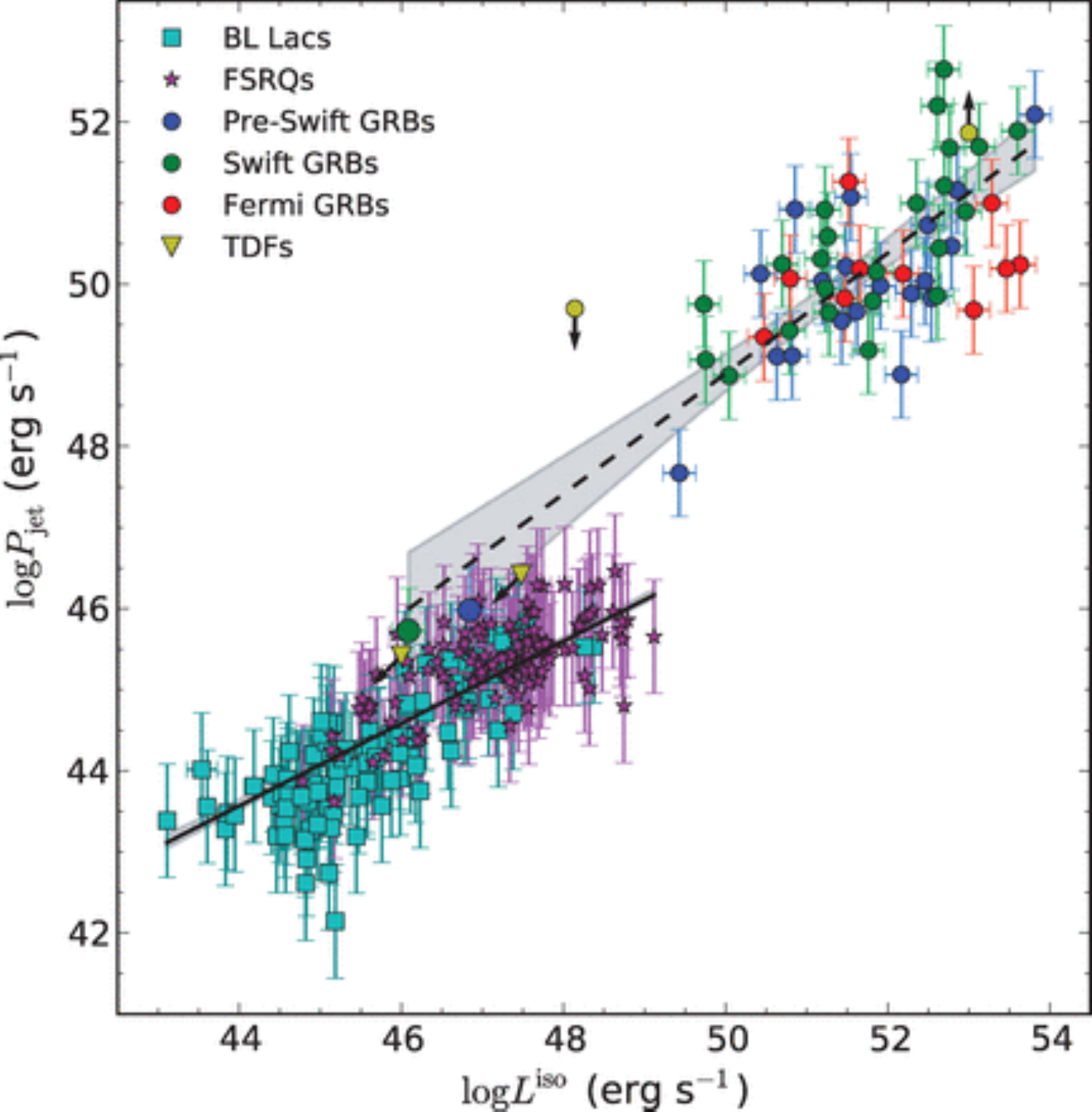}
\includegraphics[height=8.0cm]{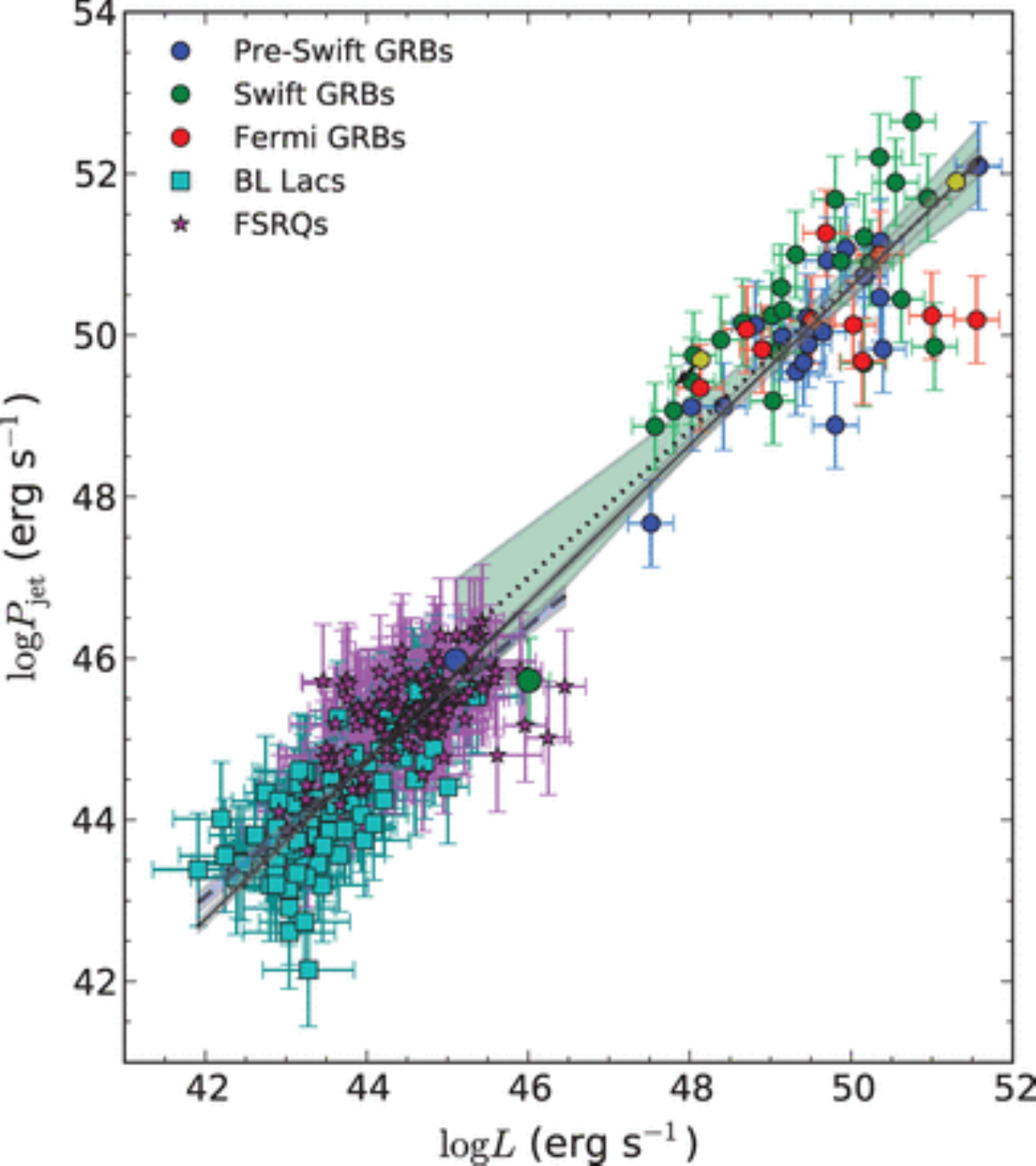}
\end{tabular}
\caption{Jet power as a function of luminosity for blazars and GRBs.  In the left panel the luminosities are  reported as observed, i.e. computed isotropically from the observed fluxes; in the right panel the luminosities are corrected for the aperture of the jet: the correlation improves significantly with respect to the left panel (from Nemmen et al. 2012).}
\label{nemmen}
\end{center}
\end{figure}

%%%%%%%%%%%%%%%%%%%%%%%%%%%%%%%%%%

\section{Blazars}
\label{blazars}

Blazars are detected from radio to very high gamma-ray frequencies, they constitute the majority of extragalactic sources detected in MeV-GeV gamma-rays (Acero et al. 2015) and are the only known cosmological  TeV emitters (Aharonian et al. 2013)\footnote{Blazars are virtually the only extragalactic  sources detected by Cherenkov telescopes, the  exceptions being the very nearby starburst galaxies M82 and NGC253 (Itoh et al. 2002; VERITAS  Collaboration 2009).}.
Their  spectral energy distributions ($\nu f_\nu$ representation) are dominated by non-thermal processes: synchrotron radiation at the lower frequencies from a population of leptons, with a spectral peak between the infrared and soft-X-ray domain, and a radiation component at high energies (hard-X- and gamma-rays) that can be due either to  inverse Compton scattering of synchrotron photons or external photons off leptons or  to proton-synchrotron
emission, $\pi_0$ decay photons, synchrotron, and Compton emission
from secondary decay products of charged pions, and the
output from pair cascades initiated by these high-energy emissions
intrinsically absorbed by photon-photon pair production (see B{\"o}ttcher et al. 2013, and  Falomo, Pian, \& Treves 2014 for reviews).   

Occasionally, the synchrotron peak can reach frequencies larger than 100 keV during outbursts,  as often observed  in the BL Lac Mkn~501   (Pian et al.  1998; Furniss et al. 2015), and, less conspicuously, in other blazars (Giommi, Padovani, \& Perlman 2000; Costamante et al. 2001).  During these  hard X-ray outbursts, the source is usually also in a high TeV state, although the radiation at these energies is dramatically affected by the large Klein-Nishina cross-section, if due to synchrotron self-Compton scattering.    Sources with this extreme behavior are the tracers of the most energetic radiating particles, so that their identification represents an effective investigation tool of the most powerful and efficient acceleration mechanisms in astrophysics.   Optimal methods to select ``extreme synchrotron'' blazar candidates include TeV detection (for the nearest sources, that are less affected by extragalactic background suppression, Costamante 2013) and a characteristic multi-wavelength spectral shape (Bonnoli et al. 2015).  Interestingly, not all extreme-synchrotron candidates actually display extremely high peak energies during outbursts, the most remarkable case being Mkn~421:  a very close spectral analogue of Mkn~501, its  synchrotron peak energy never exceeds 10-15 keV during the major historical X-ray outbursts (Pian et al. 2014, and references therein, see Figure 2), which suggests the action of an ``inhibiting'' parameter (perhaps an external photon field that is  more significant than in Mkn~501).

%%%%%%%%%%%    FIGURE 2:  Mkn421   %%%%%%%%%%%

\begin{figure} 
\begin{center}
%\hspace{5.0cm}
\includegraphics[height=15.0cm]{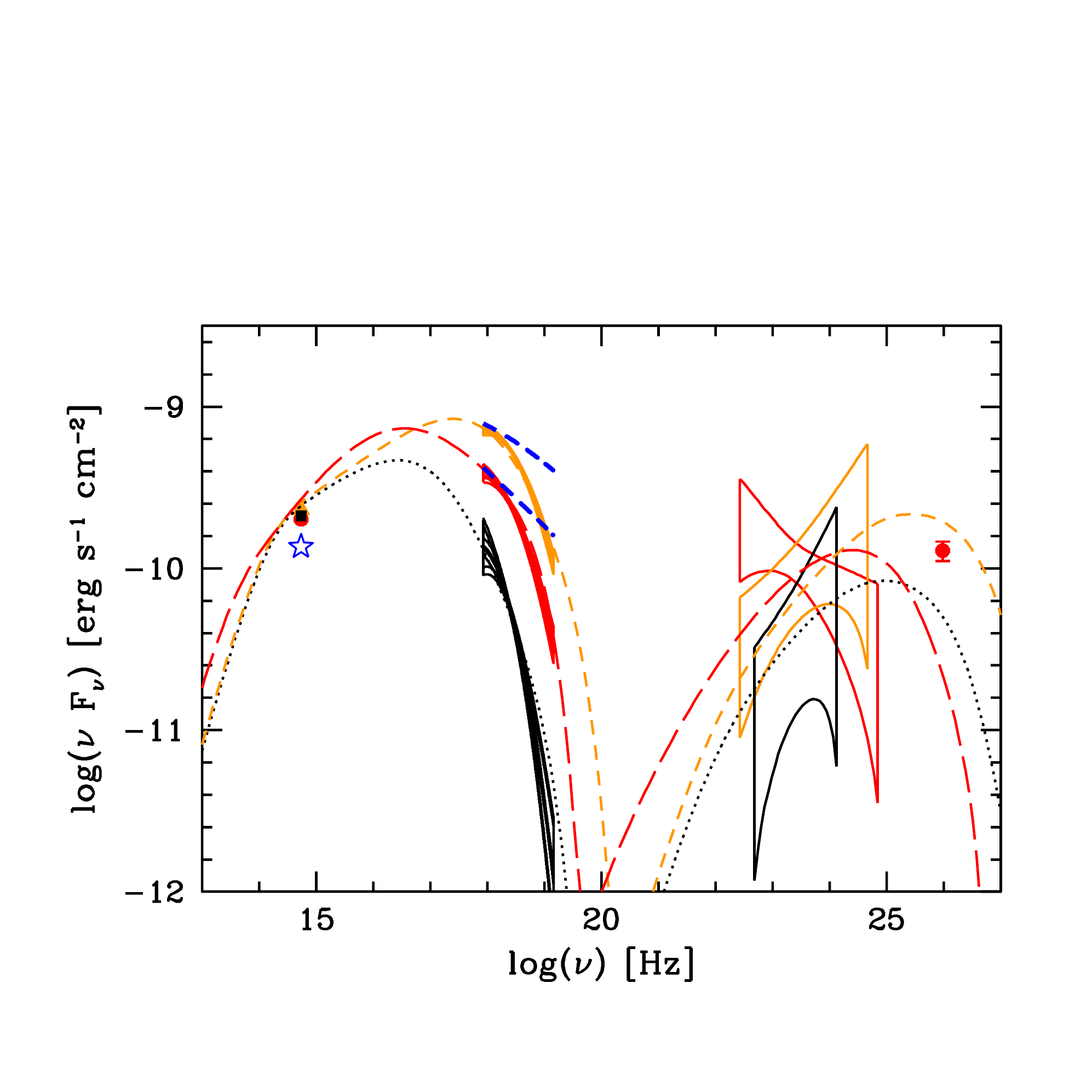}
\caption{Spectral energy distributions of Mkn~421  at the average UT epochs of 16.1--16.5 (red), 17.1--17.4 (orange), and 19.0--19.5 (black) April 2013   from simultaneous {\it INTEGRAL} IBIS/ISGRI, JEM-X and OMC, and {\it Fermi}-LAT data.  The optical data were corrected for Galactic absorption and for the contamination by galaxies in the field.  The 1-$\sigma$ error contours of the  joint JEM-X and IBIS/ISGRI spectra and LAT spectra are  reported.  The TeV point from VERITAS (red), taken on April 16.3, is also shown.   The models (long dash:  April 16, short dash: April 17, dot:  April 19) include a synchrotron component at the lower energies, produced in a single emitting zone, and a synchrotron self-Compton scattering component at  higher energies.  For comparison, the JEM-X and IBIS/ISGRI quiescent and active state spectra of June 2006 are reported as thick dashed blue lines.  The average optical flux at the same epoch is shown as a  blue star (see Pian et al. 2014 and Lichti et al. 2008).}
\label{mkn421}
\end{center}
\end{figure}

%%%%%%%%%%%%%%%%%%%%%%%%%%%%%%%%%%

A  relationship links the total luminosity and the location of the characteristic spectral peaks in blazars, in the sense that blazars of higher luminosity tend to have these peaks at lower frequencies (with exceptions, Padovani, Giommi, \& Rau 2012; Arsioli et al. 2015).    This ``blazar sequence'' (Fossati et al. 1998) implies an anti-correlation between the radiation energy density and the particles break energy, that indicates a rough constant cooling rate for all sources near peak energy (Ghisellini et al. 1998; Ghisellini 1999).    A viable interpretation consists in the different role played by the external photon fields in the cooling, more luminous sources having more prominent accretion disk components and more intense broad emission lines, that make  cooling of accelerated particles via inverse Compton scattering more efficient.
However, in individual sources the opposite behavior is observed during outbursts: the luminosity and the peak energy both rise  and decrease simultaneously in a correlated way during the outburst, likely because the equilibrium configuration is temporarily lost (see e.g. Tavecchio et al. 2000).

\section{Gamma-ray bursts}
\label{grbs}

Nearly twenty years after the beginning of the ``afterglow era'', started with the rapid and accurate localizations of  the {\it BeppoSAX} satellite (Costa et al. 1997), over 1000 GRBs have been localized at gamma- and X-rays by various space missions - the majority with the {\it Swift} satellite (Gehrels et al. 2004) - and have well observed multi-wavelength counterparts.   These witness the presence and behavior of jets, whose geometry, structure, composition and dynamics are 
still matter of investigation.  The biggest crucible related to GRBs however is the nature of their progenitors and the jet powering mechanism.  While  short GRBs\footnote{The  majority of observed GRBs have durations longer than $\sim$2 s (Kouveliotou et al. 1993) and  are thus defined as long GRBs, as opposed to short or sub-second GRBs.}  seem to be compatible with the merger of a coalescing binary compact star system, whose indirect evidence may be represented by a nucleo-synthetic near-infrared signal (Tanvir et al. 2013; Berger, Fong, \& Chornock  2013), long GRBs 
are connected with supernovae: with two exceptions, that raised however considerable debate (Della Valle et al. 2006a; Fynbo et al. 2006; Gal-Yam et al. 2006;  Gehrels et al. 2006; Ofek et al. 2007; McBreen et al. 2008; Jin et al. 2015), all GRBs at $z \simlt 0.2$ are associated with supernovae that were classified as type Ic (i.e. core-collapse supernovae with stripped hydrogen and helium envelopes) based on the unambiguous identification of  typical atomic species in their ejecta (e.g., Galama et al. 1998; Hjorth et al. 2003; Malesani et al. 2004; Modjaz et al. 2006; Pian et al. 2006; Bufano et al. 2012; D'Elia et al. 2015).  At higher redshifts, the identification of spectroscopic features is made more arduous by the contamination of the host galaxy and afterglow, but a general spectral resemblance with GRB-supernovae at lower redshifts is found, or, lacking a reliable spectrum, the SN presence is signalled by the rebrightening of the light curve at around 10-15 rest-frame days after GRB explosion (Della Valle et al. 2003;2006b; Fynbo et al. 2004; Soderberg et al. 2005;2006; Bersier et al. 2006; Berger et al. 2011;  Cano et al. 2014; Melandri 2012;2014).    
The bolometric light curves  of the two nearest known GRB-SNe, SN1999bw and SN2006aj and of  a  regular (i.e. not accompanied by a high energy transient) Ic supernova, SN1994I, are shown in Figure 3.  

%%%%%%%%    FIGURE 3:  GRB-SNe LCs    %%%%%%%%%%

\begin{figure} 
\begin{center}
%\hspace{5.0cm}
\includegraphics[height=15.0cm]{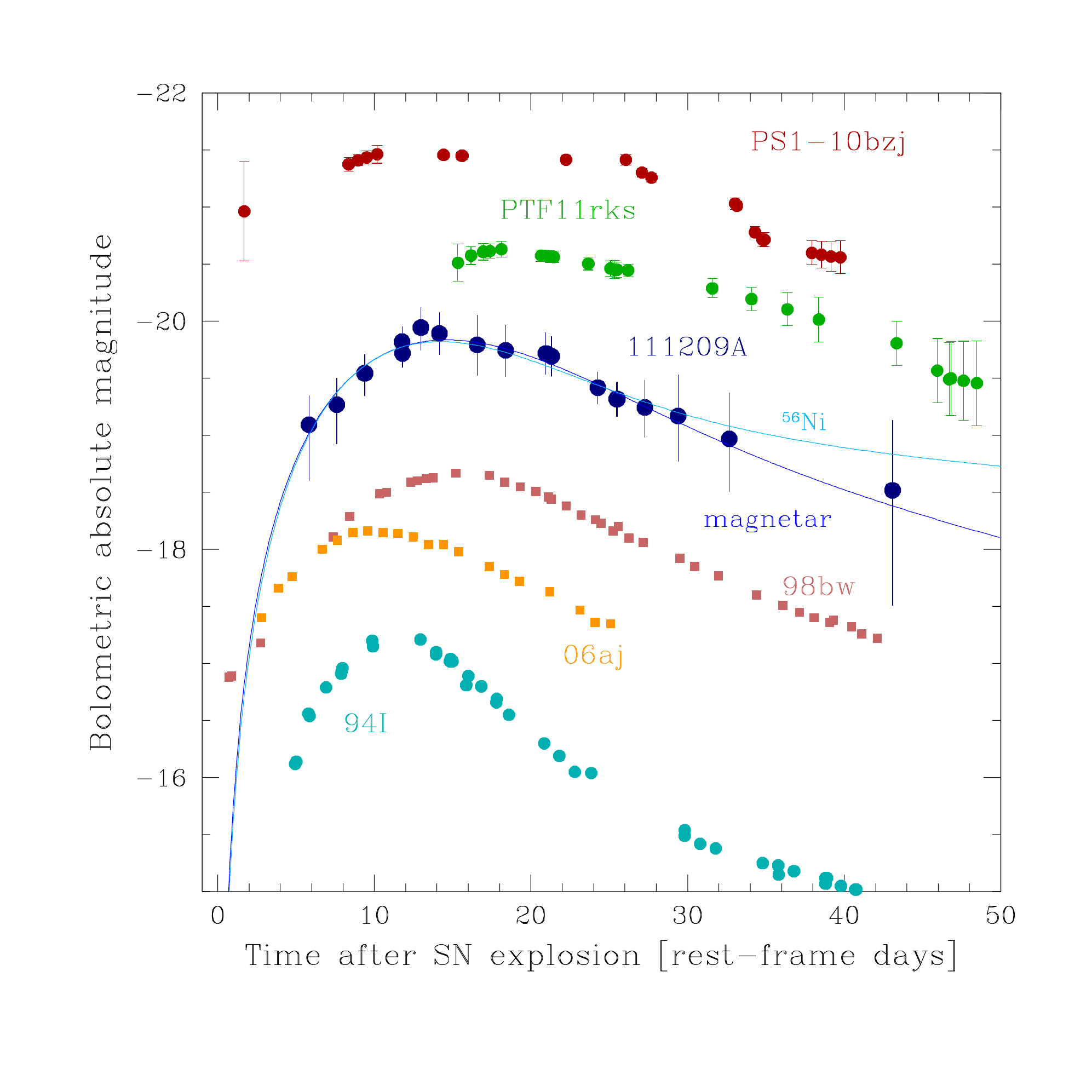}
\caption{Bolometric light curves of  the ``classical'' GRB supernovae
GRB980425/SN 1998bw and XRF 060218/SN 2006aj, the standard type Ic SN1994I, the super-luminous supernovae PTF11rks and PS1-10bzj (among the fastest declining super-luminous supernovae known so far) and the SN 2011kl associated with the ultra-long GRB111209A  (observed duration of $\sim$10000 seconds).  SN2011kl has an intermediate maximum luminosity between the GRB supernovae and the super-luminous supernovae.    Solid lines show the best-fitting
synthetic light curves computed with a magnetar injection model (dark blue) and $^{56}$Ni powering (light blue).  See Greiner et al. (2015) for more details and references.}
\label{greiner}
\end{center}
\end{figure}

%%%%%%%%%%%%%%%%%%%%%%%%%%%%%%%%%%

The most remarkable aspect  of GRB-supernovae is their kinetic energies, that are of the order of $\sim 10^{52}$ erg or higher i.e. a factor 10 or more larger than those of regular core-collapse supernovae (Mazzali et al. 2006a).  
When a correction for jet collimation is applied to the gamma-ray energy output of the GRB, this is a small fraction (less than 10\%) of the supernova kinetic energy, indicating that the energetically dominant player in the phenomenon is the supernova, not the GRB (Woosley \& Bloom 2006; Mazzali et al. 2014).  This has led to the scenario in which a rapidly rotating newly formed neutron star with a large magnetic field, a magnetar, may be responsible for GRBs, by powering the supernova with its rotational energy (which is indeed of the order of $\sim 10^{52}$ erg for a millisecond proto-neutron star) and producing a  relativistic outflow along its rotation axis, where energy is radiated away  via magnetic dipole mechanism in a time compatible with a GRB duration if the field is of the order of $\sim 10^{14}$ Gauss, as typically seen  in a magnetar (Usov 1992).  This picture has gathered  impulse (Mazzali et al. 2006b; Kasen \& Bildsten 2010; Metzger et al. 2011;2015) as an alternative to the collapsar, that envisages GRBs originating from accretion disks promptly formed around black holes that result from massive stars core collapse (MacFadyen \& Woosley 1999).

Among the findings of {\it Swift} is the detection of an observationally  rare class of GRBs,  whose duration of about 10000 seconds exceeds significantly the average duration of long GRBs.   The afterglow of one of these  ``ultra-long''  GRBs, GRB111209A, was thoroughly studied at X-ray and optical wavelengths (Gendre et al. 2013; Stratta et al. 2013; Levan  et al. 2014), and the regular and intensive optical/near-infrared monitoring with the GROND instrument on the ESO 2.2m  telescope revealed the presence of a supernova, dubbed SN2011kl  (Greiner et al. 2015).   SN2011kl does not resemble any of the GRB-supernovae previously detected, being more luminous and very poor in metals.  Its luminosity places  it at halfway between ``classical''  GRB-supernovae and the so called super-luminous supernovae (see Figure 3), a class of supernovae recently discovered, very massive and luminous, a fraction of which may  be related  to pair instability in the stellar nucleus (Gal-Yam 2012).    

The high amount of radioactive nickel ($^{56}$Ni)  that is formally necessary to account for the light curve of SN2011kl ($\sim$1 M$_\odot$) is inconsistent with the low metal content inferred from the optical spectrum, suggesting that an extra component is responsible for powering the supernova.  Since the accretion rates implied in a collapsar scenario are incompatible with the GRB duration  (the progenitor core mass would exceed many hundreds of solar masses), the magnetar alternative seems viable and even more cogent  than it is for the supernovae accompanying ``regular''  long GRB.  The fact that SN2011kl is more luminous than the other GRB-supernovae and thus more similar to super-luminous supernovae, is an added element in favour of magnetar, that was proposed as the engine of super-luminous supernovae  (Woosley 2010).

\section{Conclusions}
\label{conclusion}

Although it is not yet clear what ultimately causes and triggers  blazar multi-wavelength outbursts, the strong correlation between broad emission line luminosity and jet luminosity indicates that the control parameter for both is accretion (Ghisellini et al. 1998; Sbarrato et al. 2012).   Accretion may also play a crucial role in GRBs if a fraction or all of them are powered by rotating black holes, as in the collapsar model.  The biggest open problem would then consist in  understanding how physical properties scale with central black hole mass  in the two phenomena, considered that energy dissipation appears to correlate with jet power in a very similar way  (Nemmen et al. 2012). 
 
A further complication may descend from the detection of at least one case of a GRB-supernova association, GRB111209A/SN2011kl,  where a magnetar scenario (rapidly rotating, highly magnetized proto-neutron star), rather than a collapsar (accreting black hole), seems not only viable but preferred.  The fact that both the GRB and the supernova belong to peculiar types (the GRB is ultra-long  and the supernova is a factor of 3 more luminous than previously detected GRB supernovae  and a factor  of 3 less luminous than fast-declining super-luminous supernovae) compounds the issue and points to a novel aspect of GRB-supernova diversity.    

More observations in the form of dedicated campaigns and strategically designed multi-wavelength monitoring are necessary to clarify these issues.  Specifically, these include mapping of blazar jets with correlated rapid X-ray and TeV variability (see e.g., Albert et al. 2007), the latter becoming  possible within the next decade thanks to the sensitive next generation Cherenkov Telescope Array, and accurate optical and high energy observations of ultra-long GRBs.

A very ambitious goal is the construction of a unifying scenario for long GRBs and supernovae, analogous to the one that links FR I and FR II radio-galaxies to BL Lac objects and Flat-Spectrum Radio Quasars, respectively.  Since GRBs have jets of a few degrees aperture, if all the long ones, or a large fraction thereof, are related to energetic supernovae, we should be observing -- within a given cosmological volume -- many more energetic supernovae than GRBs, because the vast majority of GRB jets would be misaligned with respect to our line of sight.  The present sensitive all-sky high cadence optical surveys are detecting numerous energetic  Ic supernovae (e.g. Mazzali  et al. 2013); however, establishing the presence of an accompanying misaligned GRB is difficult.  Multi-wavelength relativistic jet models (van Eerten, Zhang, \& MacFadyen 2010) predict that a rebrightening in the radio light curve of a GRB afterglow should occur months or years after explosion as the jetted blast wave isotropizes.  Lacking a detection in gamma- or X-rays, a misaligned GRB jet should therefore become detectable as enhanced radio emission from the location of its energetic supernova progenitor.  This requires however a very sensitive radio array and  long uninterrupted exposures of a sizable sample of  high kinetic energy supernovae.  The newly deployed and developing LOFAR experiment may be suited for this investigation. In the case of short GRBs, an analogous search for transients at low-frequency radio wavelengths (Nakar \& Piran 2011) may reveal the counterparts of double neutron star mergers even in the absence of a detected short GRB.  In addition, it would contribute to validate any gravitational radiation signal emitted right before the merger.

\acknowledgments  I would like to thank Gustavo Romero and Gabriela Vila for organizing a most stimulating and successful conference, and the Observatory of La Plata for hospitality and support.  Financial support from the Italian Ministry of Education, University and Research and Scuola Normale Superiore of Pisa is also acknowledged.


\begin{references}

\reference Acero, F., Ackermann, M.,  Ajello, M., et al.\ 2015, \apjs, 218, 23 
\reference Aharonian, F., Essey, W., Kusenko, A., \& Prosekin, A.\ 2013, \prd, 87, 063002 
\reference Albert, J., Aliu, E.,  Anderhub, H., et al.\ 2007, \apj, 669, 862 
\reference Amati, L., Guidorzi, C.,  Frontera, F., et al.\ 2008, \mnras, 391, 577 
\reference Arsioli, B., Fraga, B., Giommi, P.,  Padovani, P., \& Marrese, P.M. 2015,  A\&A, 579, A34
\reference Berger, E., Chornock,  R., Holmes, T.~R., et al.\ 2011, \apj, 743, 204 
\reference  Berger, E., Fong, W.,  \& Chornock, R.\ 2013, \apjl, 774, L23 
\reference Bersier, D., Fruchter,  A.~S., Strolger, L.-G., et al.\ 2006, \apj, 643, 284 
\reference Beuermann, K., Hessman, F.~V., Reinsch, K., et al.\ 1999, \aap, 352, L26 
\reference B{\"o}ttcher, M.,  Reimer, A., Sweeney, K., \& Prakash, A.\ 2013, \apj, 768, 54 
\reference Bonnoli, G., Tavecchio, F., Ghisellini, G., \& Sbarrato, T.\ 2015, \mnras, 451, 611 
\reference Bufano, F., Pian, E., Sollerman, J., et al.\ 2012, \apj, 753, 67 
\reference Cano, Z., de Ugarte Postigo, A., Pozanenko, A., et al.\ 2014, \aap, 568, A19 
\reference Costa, E., Frontera, F.,  Heise, J., et al.\ 1997, \nat, 387, 783 
\reference Costamante, L., Ghisellini, G., Giommi, P., et al.\ 2001, \aap, 371, 512 
\reference Costamante, L.\ 2013,  International Journal of Modern Physics D, 22, 1330025 
\reference D'Elia, V., Pian, E., Melandri, A., et al.\ 2015, \aap, 577, A116 
\reference Della Valle, M., Malesani, D., Benetti, S., et al.\ 2003, \aap, 406, L33
\reference Della Valle, M.,  Chincarini, G., Panagia, N., et al.\ 2006a, \nat, 444, 1050 
\reference Della Valle, M., Malesani, D., Bloom, J.~S., et al.\ 2006b, \apjl, 642, L103 
\reference Falomo, R., Pian, E., \& Treves, A.\ 2014, \aapr, 22, 73 
\reference Fender, R., \& Belloni, T.\ 2004, \araa, 42, 317 
\reference Fossati, G., Maraschi,  L., Celotti, A., Comastri, A., \& Ghisellini, G.\ 1998, \mnras, 299, 433 
\reference Frail, D.~A., Kulkarni,  S.~R., Nicastro, L., Feroci, M., \& Taylor, G.~B.\ 1997, \nat, 389, 261 
\reference Frail, D.~A., Kulkarni,  S.~R., Sari, R., et al.\ 2001, \apjl, 562, L55 
\reference Furniss, A., Noda, K.,  Boggs, S., et al.\ 2015, \apj, 812, 65 
\reference Fynbo, J.~P.~U., Sollerman, J., Hjorth, J., et al.\ 2004, \apj, 609, 962 
\reference Fynbo, J.~P.~U., Watson, D., Th{\"o}ne, C.~C., et al.\ 2006, \nat, 444, 1047 
\reference Galama, T.~J., Vreeswijk, P.~M., van Paradijs, J., et al.\ 1998, \nat, 395, 670 
\reference Gal-Yam, A., Fox,  D.~B., Price, P.~A., et al.\ 2006, \nat, 444, 1053 
\reference  Gal-Yam, A.\ 2012, Science, 337, 927 
\reference Gehrels, N., Chincarini, G., Giommi, P., et al.\ 2004, \apj, 611, 1005 
\reference Gehrels, N., Norris, J.~P., Barthelmy, S.~D., et al.\ 2006, \nat, 444, 1044 
\reference Gendre, B., Stratta, G.,  Atteia, J.~L., et al.\ 2013, \apj, 766, 30 
\reference Ghisellini, G., Celotti, A., Fossati, G., Maraschi, L.,  \& Comastri, A.\ 1998, \mnras, 301, 451 
\reference Ghisellini, G.\ 1999,  Astroparticle Physics, 11, 11 
\reference Giommi, P., Padovani,  P., \& Perlman, E.\ 2000, \mnras, 317, 743 
\reference Greiner, J., Mazzali, P.~A., Kann, D.~A., et al.\ 2015, \nat, 523, 189 
\reference Grupe, D., Brown, P.~J.,  Cummings, J., et al.\ 2006, \apj, 645, 464 
\reference Harrison, F.~A., Bloom, J.~S., Frail, D.~A., et al.\ 1999, \apjl, 523, L121 
\reference Hjorth, J., Sollerman,  J., M{\o}ller, P., et al.\ 2003, \nat, 423, 847 
\reference Israel, G.~L., Marconi, G., Covino, S., et al.\ 1999, \aap, 348, L5 
\reference Itoh, C., Enomoto, R., Yanagita, S., et al.\ 2002, \aap, 396, L1 
\reference Jin, Z.-P., Li, X., Cano,  Z., et al.\ 2015, \apjl, 811, L22 
\reference Jorstad, S.~G., Marscher, A.~P., Smith, P.~S., et al.\ 2013, \apj, 773, 147
\reference Kasen, D., \& Bildsten, L.\ 2010, \apj, 717, 245 
\reference Kouveliotou, C.,  Meegan, C.~A., Fishman, G.~J., et al.\ 1993, \apjl, 413, L101 
\reference Levan, A.~J., Tanvir,  N.~R., Starling, R.~L.~C., et al.\ 2014, \apj, 781, 13 
\reference Lichti, G.~G., Bottacini, E., Ajello, M., et al.\ 2008, \aap, 486, 721 
\reference MacFadyen, A.~I., \& Woosley, S.~E.\ 1999, \apj, 524, 262 
\reference Malesani, D., Tagliaferri, G., Chincarini, G., et al.\ 2004, ApJ, 609, L5 
\reference Maraschi, L.,  Ghisellini, G., \& Celotti, A.\ 1992, \apjl, 397, L5 
\reference Mazzali, P.~A., Deng,  J., Pian, E., et al.\ 2006a, \apj, 645, 1323 
\reference Mazzali, P.~A., Deng,  J., Nomoto, K., et al.\ 2006b, \nat, 442, 1018 
\reference Mazzali, P.~A., Walker, E.~S., Pian, E., et al.\ 2013, \mnras, 432, 2463 
\reference Mazzali, P.~A., McFadyen, A.~I., Woosley, S.~E., Pian, E., 
\& Tanaka, M.\ 2014, \mnras, 443, 67 
\reference  McBreen, B.\ 1979, \aap, 71, L19 
\reference McBreen, S., Foley, S.,  Watson, D., et al.\ 2008, \apjl, 677, L85 
\reference Melandri, A., Pian, E., Ferrero, P., et al.\ 2012, \aap, 547, A82 
\reference Melandri, A., Pian, E., D'Elia, V., et al.\ 2014, \aap, 567, A29 
\reference  Mesler, R.~A.,  Pihlstr{\"o}m, Y.~M., Taylor, G.~B., \& Granot, J.\ 2012, \apj, 759, 4 
\reference M{\'e}sz{\'a}ros, P.\ 2002, \araa, 40, 137 
\reference Metzger, B.~D., Giannios, D., Thompson, T.~A., Bucciantini, N., 
\& Quataert, E.\ 2011, \mnras, 413, 2031 
\reference  Metzger, B.~D.,  Margalit, B., Kasen, D., \& Quataert, E.\ 2015, \mnras, 454, 3311 
\reference Mirabel, I.~F., \& Rodr{\'{\i}}guez, L.~F.\ 1999, \araa, 37, 409 
\reference Modjaz, M., Stanek,  K.~Z., Garnavich, P.~M., et al.\ 2006, \apjl, 645, L21
\reference Nakar, E., \& Piran, T.\ 2011, \nat, 478, 82  
\reference Nemmen, R.S.,  Georganopoulos,  M., Guiriec, S.,  Meyer,  E.T., \&  Sambruna, R.M. 2012, Science, 338, 1445
\reference Ofek, E.~O., Cenko, S.~B.,  Gal-Yam, A., et al.\ 2007, \apj, 662, 1129 
\reference Padovani, P., Giommi, P., \& Rau, A. 2012, MNRAS, 422, L48
\reference Panaitescu, A., \& Kumar, P.\ 2002, \apj, 571, 779 
\reference Pian, E., Vacanti, G., Tagliaferri, G., et al.\ 1998, \apjl, 492, L17 
\reference Pian, E., Mazzali, P.~A.,  Masetti, N., et al.\ 2006, \nat, 442, 1011 
\reference Pian, E., T{\"u}rler, M., Fiocchi, M., et al.\ 2014, \aap, 570, A77 
\reference Piran, T.\ 2004, Reviews of  Modern Physics, 76, 1143 
\reference Rhoads, J.~E.\ 1999, \apj, 525,  737 
\reference Sbarrato, T.,  Ghisellini, G., Maraschi, L., \& Colpi, M.\ 2012, \mnras, 421, 1764 
\reference Soderberg, A.~M.,  Kulkarni, S.~R., Fox, D.~B., et al.\ 2005, \apj, 627, 877 
\reference Soderberg, A.~M.,  Kulkarni, S.~R., Price, P.~A., et al.\ 2006, \apj, 636, 391 
\reference Stanek, K.~Z., Garnavich, P.~M., Kaluzny, J., Pych, W., \& Thompson, I.\ 1999, \apjl, 522, L39 
\reference Stratta, G., Gendre,  B., Atteia, J.~L., et al.\ 2013, \apj, 779, 66 
\reference Tanvir, N.~R., Levan,  A.~J., Fruchter, A.~S., et al.\ 2013, \nat, 500, 547 
\reference Tavecchio, F.,  Maraschi, L., Pian, E., et al.\ 2001, \apj, 554, 725 
\reference Urry, C.~M., \& Padovani, P.\ 1995, \pasp, 107, 803 
\reference Usov, V.~V.\ 1992, \nat, 357, 472 
\reference van Eerten, H., Zhang, W., \& MacFadyen, A.\ 2010, \apj, 722, 235 
\reference VERITAS  Collaboration, Acciari, V.~A., Aliu, E., et al.\ 2009, \nat, 462, 770 
\reference Woosley, S.~E., \& Bloom, J.~S.\ 2006, \araa, 44, 507 
\reference Woosley, S.~E.\ 2010, \apjl,  719, L204 
\end{references}
\end{document}